\documentclass{raa}           
\usepackage{graphicx,times}
\usepackage{natbib}
\usepackage{amssymb,amsmath}

\newcommand{\ha}{\hbox{H$\alpha$}}
\newcommand{\hb}{\hbox{H$\beta$}}
\newcommand{\hd}{\hbox{H$\delta$}}
\newcommand{\hg}{\hbox{H$\gamma$}}
\newcommand{\he}{\hbox{H$\epsilon$}}
\newcommand{\oi}{\hbox{[O\,{\sc i}]}}
\newcommand{\oii}{\hbox{[O\,{\sc ii}]}}
\newcommand{\nii}{\hbox{[N\,{\sc ii}]}}
\newcommand{\sii}{\hbox{[S\,{\sc ii}]}}
\newcommand{\oiii}{\hbox{[O\,{\sc iii}]}}
\newcommand{\neiii}{\hbox{[Ne\,{\sc iii}]}}
\newcommand{\hei}{\hbox{[He\,{\sc i}]}}
\newcommand{\dfn}{\hbox{$D_{\rm n}$(4000)}}
\newcommand{\hda}{\hbox{H$\delta_{\rm A}$}}
\newcommand{\zoh}{\hbox{$12\,+\,{\rm log(O/H)}$}}
\newcommand{\etal}{\hbox{et\thinspace al.\ }}
\newcommand{\mum} {\hbox{$\mu{\rm m}$}}
\newcommand{\Zsun}{\hbox{${\rm Z}_{\odot}$}}

\begin{document}

   \title{Spectral Synthesis
via Mean Field approach Independent Component Analysis
}

   \volnopage{Vol.0 (200x) No.0, 000--000}    
   \setcounter{page}{1}         

   \author{Ning Hu \and Shan-Shan Su \and Xu Kong}

   \institute{
        CAS Key Laboratory for Researches in Galaxies and Cosmology, 
        Department of Astronomy, University of Science and Technology of China,
        Hefei, Anhui 230026, China 
        {\it huning@mail.ustc.edu.cn; xkong@ustc.edu.cn}\\
          }

  \date{Received 2015 ? ?; accepted 2015 ? ?}

\abstract{ 
In this paper, we apply a new statistical analysis technique, Mean Field
approach to Bayesian Independent Component Analysis (MF-ICA), on galaxy
spectral analysis.  This algorithm can compress the stellar spectral 
library into a few Independent Components (ICs), and galaxy spectrum can 
be reconstructed by these ICs. Comparing to other algorithms
which decompose a galaxy spectrum into a combination of several simple 
stellar populations, MF-ICA approach offers a large improvement in the 
efficiency. 
To check the reliability of this spectral analysis method, three different
 methods are used: (1) parameter-recover for simulated galaxies, 
(2) comparison with parameters estimated by other methods, and 
(3) consistency test of parameters from the Sloan Digital Sky Survey 
galaxies. 
We find that our MF-ICA method not only can fit the observed galaxy spectra 
efficiently, but also can recover the physical parameters of galaxies 
accurately.
We also apply our spectral analysis method to the DEEP2 spectroscopic data, 
and find it can provide excellent fitting for those low signal-to-noise 
spectra.
}

   \authorrunning{N. Hu, S.-S. Su  \& X. Kong } 
   \titlerunning{Spectral Synthesis via MF-ICA} 
   \maketitle

%
\section{Introduction}          \label{sect:intro}

Spectrum contains plentiful information about the properties of
galaxy \citep{2014ChA&A..38..427K}. Finding a way to analyze the spectra 
of observed galaxies and
determine the parameters of large sample of galaxies would not only
help us to investigate galaxy formation and evolution, but also allow
us to derive cosmological information from a large number of galaxies
\citep{2013ARA&A..51..393C}.
Many methods, based on the used features, were devised to measure 
and understand the physical parameters of galaxies, whether by using the 
spectral indices \citep{wo94} or emission features 
\citep{ke01,2014MNRAS.444L..49S}, or by fitting
full spectrum \citep{tr04,cf05,oc06,to07,2013arXiv1305.2984L}. 
Due to the abundance of high-quality
galaxy spectra, two different population synthesis approaches have been 
commonly used to study the stellar contents of galaxy.  Empirical population 
synthesis method \citep{fa72,bi88,cf01,2003A&A...403..877K} is based on 
modelling galaxies as mixture 
of several observed spectra of stars or star clusters. However, this method 
do not consider the stellar evolution and is limited by the observed 
stellar/cluster
spectral library.  Recently, a more direct approach so-called evolutionary 
population synthesis \citep{va99,gi00,bc03,2005MNRAS.362..799M,
2015MNRAS.452.1068C} has been widely used.  
In this approach, the spectra of stellar populations are modeled 
by combining stellar evolution tracks, SSP library and star formation 
histories (SFHs).  Up to now, a popular simple stellar populations 
(SSPs) library was provided by the isochrone synthesis technique (BC03). 
Several group have selected a few SSPs from this library as templates 
to fit the observed galaxy spectra \citep{tr04,cf05}.

However, the advent of large-area spectroscopic surveys, such as
the Third Sloan Digital Sky Survey \citep{ei11}, the Deep Extragalactic
Evolutionary Probe 3 (DEEP3) Galaxy Redshift Survey \citep{co11}, and the 
Large sky Area Multi-Object fiber Spectroscopic Telescope \citep{cu12}, 
will be providing oceans of data, thus 
the development of fast and automated extraction methods are required. 
We note 
that statistical analysis techniques have been commonly implemented. 
For example, 
Richards et al. (2009) utilized the diffusion $k$-means method to draw 
several
prototype spectra from SSP database as input templates of the spectral
synthesis software STARLIGHT \citep{cf05}.  Nolan et al. (2006) applied a 
data-driven Bayesian approach to the spectra of early-type galaxies.
Another blind source separation (BSS) technique applied to spectra is 
principal component analysis (PCA, \citealt{mitt90}, 
\citealt{2001MNRAS.323.1035K}; \citealt{yip04}), but the interpretation 
of the individual component spectra seems rarely illuminating.
Here, we explore a new statistical multivariate
data processing technique, independent component analysis (ICA), in our
spectral analysis.  This technique has been implemented in the Cosmic 
Microwave Background studies \citep{ma07} and the analysis of spectra 
\citep{lu06,all13}, 
however, the Ensemble Learning ICA (EL-ICA, also known as naive mean 
field or NMF) method used in Lu et al. (2006) is known to fail in some
 circumstances (e.g. low signal-to-noise spectra) \citep{hs01}, and 
Allen et al. (2013) applied to only emission-line galaxies.  
For the sake of non-negative values in the galaxy spectral analysis, 
we adopt a new ICA
algorithm, mean field approach independent component analysis (MF-ICA),
which can constrain the sources and the mixing matrix to be non-negative
with a more efficient and more reliable algorithm.

The paper is structured as follows. In Section 2, we introduce
the MF-ICA method, and derive a few templates from evolutionary
 population models
of Charlot \& Bruzual (2007) to analyze the spectra of galaxy.
In Section 3, the simulated galaxy spectra are used to analyze the
reliability of the MF-ICA method.  In Section 4, we analyze the SDSS
observed galaxy spectra, compare our results with those obtained from
the MPA/JHU\footnote{http://www.mpa-garching.mpg.de/SDSS/} catalogs, to
investigate whether our synthesis results are reasonable.
In Section 5, some galaxy spectra from the DEEP2 galaxy redshift survey
are fitted by our method, and our conclusions are outlined in Section 6.

\section{Method}\label{sec:method}

\subsection{Stellar Population Models}\label{sec:ssp}

Stellar population models can be generated by several population
synthesis codes, here we adopt the 2007 version of 
Galaxev\footnote{http://www.bruzual.org} 
\citep{cb07}, which is a new version of BC03. The
CB07 models have undergone a major improvement recently with the new
stellar evolution prescriptions of Marigo \& Girardi (2007) for the
Thermally-Pulsing Asymptotic Giant Branch (TP-AGB) evolutionary phase of
low- and intermediate-mass stars.  An accurate modelling of this phase
is related to correct predicted fluxes at the 
wavelength range of $1 - 2.5$\mum\ (Charlot \& Bruzual, 2007).

The CB07 models use an empirical spectral library with a range of
wavelength (91\AA $ - 36000\mum$, $N=6917$), and spectral resolution
about 3\AA.  Moreover, CB07 contains a large sample of SSP,
which covers 221 different ages from $1.0
\times 10^5$ to $2.0 \times 10^{10}$ yr, and a wide range of initial
chemical compositions, $Z = 0.0001$, 0.0004, 0.004, 0.008, 0.02,
0.05 and 0.1 ($\Zsun=0.02$).  The observed spectrum of a galaxy can
be expressed as a combination of these individual SSPs with weights.
This SSPs database will be used to derive our templates in 
\S\,\ref{sec:ssica}.

\subsection{MF-ICA technique}

\subsubsection{Independent Component Analysis (ICA)}

ICA is a new multivariate data processing method which aims at
decomposing complex multivariate observations to a combination of a
few hidden original sources \citep{hy01}.  Comparing to
the traditional multivariate data processing methods, such as principal
component analysis (PCA) or
factor analysis, ICA is much powerful for finding the hidden sources,
even when those traditional methods fail completely.  The following
generative model of ICA shows that multivariate observations or mixed
signals $x^{i}$, i=1, 2, ..., m, are a combination of hidden sources,
i.e. Independent Components, $h_{k}$, k=1, 2, ..., n, with additive
Gaussian noise $\Gamma^{i}$, weighted by the mixing weights $w_{k}^{i}$
(m $\times$ n):
\begin{equation} 
x^{i}=\sum_{k=1}^n w_{k}^{i}h_{k} + \Gamma^{i}\ \ \ \ \ (i=1,2,...,m).
\end{equation}

In our analysis, we take the multivariate observations as the spectra of
stellar systems (e.g. SSP database), and adopt the assumption that each
spectrum $f^{i}(\lambda)$ can be expressed as a sum of several Independent
Components (ICs), $IC_{k}(\lambda)$, so the model can be written as:
\begin{equation}
f^{i}(\lambda)=\sum_{k=1}^n w_{k}^{i}IC_{k}(\lambda)
+ \Gamma^{i}(\lambda)
\end{equation}
Here, we only know the spectrum $f^{i}(\lambda)$, the unknown mixing
weights $w_{k}^{i}$, the independent components $IC_{k}(\lambda)$ and the
noise can be estimated from ICA algorithms, such as Joint Approximate
Diagonalization of Eigenmatrices \citep{cs93},
extended InfoMax \citep{bs95}, FastICA \citep{hy01}, Ensemble Learning 
ICA (EL-ICA; Miskin \& Mackay 2001), Mean Field ICA \citep{hs02} and many 
others.

\subsubsection{Mean Field approach ICA (MF-ICA)}

The ICA algorithm we adopt to our spectral analysis is MF-ICA method.
Comparing to other algorithms, MF-ICA is a Bayesian iterative algorithm
which can constraint sources and mixing matrix to be positive by
offering priors of them.  The main advantage of MF-ICA algorithm is its
implementation simplicity and generality.

In this approach, the likelihood for the parameters and sources is
defined as ${\rm P({\bf X|W,\Sigma, H})}$ given by:
\begin{equation} {\rm P{\bf
(X|W,\Sigma,H)=(det2\pi\Sigma)^{-\frac{n}{2}}e^{-\frac{1}{2}Tr( X -
WH)^{T}\Sigma^{-1}(X - WH)}}},
\label{eq:likelihood}
\end{equation}
where {\bf W} is mixing matrix, {\bf X}$=[x^{1},x^{2},...,x^{m}]^{T}$ is
mixed signals matrix, ${\bf \Sigma}$ is noise covariance matrix,
n is the number of input source signal, and $det$ is the determinant 
of the matrix. 
The likelihood of the parameters is defined as ${\rm P({\bf X|W,\Sigma})}$ 
obtained from:
\begin{equation}
{\rm P({\bf X|W,\Sigma})}=\int d{\rm {\bf H}P({\bf X|W,\Sigma,H})P({\bf H})}.
\label{eq:likeofpa}
\end{equation}
If priors on the mixing weight\ ${\rm P({\bf W})}$ 
and the sources ${\rm P({\bf H})}$ are taken into account, then the posteriors 
of sources and mixing matrix are obtained from  
${\rm P({\bf H|X ,W,\Sigma}) \propto
P({\bf X|W,\Sigma,H})P({\bf H})}$ and ${\rm P({\bf W|X,\Sigma}) \propto
P({\bf X|W,\Sigma})P({\bf W})}$, respectively. In MF-ICA method, the noise
covariance ${\bf \Sigma}$ and mixing matrix {\bf W} can be obtained from
maximum a posteriori, while sources {\bf H} can be obtained from their
posterior mean. The mean field approach can be solved by:
\begin{eqnarray}
{\rm \hat{{\bf H}}=\langle {\bf H}\rangle}
\end{eqnarray}
\begin{eqnarray}
{\rm \hat{{\bf W}}={\bf X}\langle
{\bf H}^{T}\rangle\langle {\bf HH}^{T}\rangle^{-1}}
\end{eqnarray}
\begin{eqnarray}
{\rm {\bf \Sigma}=\frac{1}{n}\langle({\bf X}-\hat{{\bf
W}}\hat{{\bf H}})({\bf X-\hat{W}\hat{H}})^{T}\rangle},
\end{eqnarray}
where ${\rm \langle\cdot\rangle = \langle\cdot\rangle_{{\bf
H|W,\Sigma,X}}}$ denotes the posterior average with respect to the
sources given the mixing matrix and noise covariance.  The solution of
MF-ICA algorithm is equal to update noise covariance (Eq. 7) and mixing
matrix (Eq. 6) alternatively, and estimate sources (Eq. 5). Thus
the optimized matrices of mixing matrix $\hat{{\rm {\bf W}}}$, noise
covariance ${\rm {\bf \Sigma}}$, and sources $\hat{{\rm {\bf H}}}$
can be derived from this iterative method.  More details about the
MF-ICA method can be found in H{\o}jen-S{\o}rensen et al. (2002) and
the available {\sc Matlab} toolbox (http://mole.imm.dtu.dk/toolbox/ica).

Through the Bayesian inference of the mixing matrix and sources, their
priors can be constrained to be non-negative, which will be useful
in observed galaxy spectrum processing, since the spectra parameters
should not be negative.  Although EL-ICA method has been implemented in
 galaxy spectral
analysis \citep{lu06}, here we adopt the MF-ICA method, which relies
on advanced mean field approaches: linear response theory and an adaptive
version mean-field approach. H{\o}jen-S{\o}rensen et al. (2001, 2002) have
investigated both MF-ICA and EL-ICA methods, they concluded that comparing
to the EL-ICA method, the advanced mean field approaches can recover the
correct sources even when the ensemble learning theory fails, and the
convergence rate of MF-ICA method is found to be faster.  The comparison
of these two ICA methods will be described in \S\,\ref{sec:com}.

\subsubsection{SSPs Spectral Analysis using MF-ICA}\label{sec:ssica}

Through the multivariate data processing technique, we expect to derive
a minimal number of non-negative templates, which can represent the
spectra of galaxy with minimal loss of information.  Here, we adopt the
MF-ICA algorithm to compress the spectral library of SSP from CB07
models (\S\,\ref{sec:ssp}).

The SSP database of CB07 contains 1547 spectra (\S\,\ref{sec:ssp}),
each spectrum was first truncated to the high resolution wavelength
range of $3322 - 9200$\AA, to match that of SDSS spectrograph.
In EL-ICA method, the number of sources (i.e. ICs) should be same as
the number of mixed signals.  Therefore, Lu et al. (2006) picked up a
subsample out of the BC03 SSP database as the mixed signals matrix {\bf
X} in the EL-ICA method, and estimated 74 hidden spectra.  Finally they
choose several ICs from these hidden spectra
by the average fractional contribution to BC03 SSP database.  However,
the MF-ICA method we applied can do the dimensional reduction.  Here the
whole CB07 SSP database was set as the input mixed signals matrix {\bf
X}, then the MF-ICA method will be applied to them, and the output ICs
will be more precise.  To avoid negative values appearing in spectral
analysis, we set the priors of mixing matrix and sources as positive.

As has been mentioned above, the number of ICs can be less than
mixed signals number in MF-ICA method, thus it should be predefined.
The correct number can be determined as follows. We apply
Root-Sum-Square (RSS) method to select the proper number of ICs.
The value
of RSS between the original mixed signals (i.e. whole SSP database)
and the recovered mixed matrix can be calculated by:
\begin{equation}
{\rm RSS}=\left(\sum_{j=1}^{n} \sum_{i=1}^{m}( x^{i}_{j}-
\hat{x^{i}_{j}})^{2}\right)^{1/2} ,
\label{eq:ssr}
\end{equation}
where the
recovered mixed matrix ${\bf \hat{X}}$ is calculated from the estimated
mixing matrix and sources: $\bf \hat{X}=\bf \hat{W}\hat{H}$.  We preset the
initial number of sources as one, then increase the number and the value
of RSS will be reduced.  Repeated this process until the reduction is
no longer significant. Finally, the number of ICs can be set as {\bf 12}.

Using the number of ICs we determined, the sources can be obtained from
MF-ICA calculation. Therefore, the SSP database can be compressed
into 12 ICs, we present these 12 ICs in Figure\,4 of 
\citet{2013ApJ...778...10S}.

To confirm the reliability and quality of the ICs, we used the estimated
12 ICs to recover the 1547 CB07 SSP database as follows:
\begin{equation}
\rm f^{i}_{SSP}(\lambda)=\sum_{k=1}^{12} w_{k}^{i}{\rm
IC}_{k}(\lambda) \hspace{1.0cm} (i= 1, 2, ..., 1547),
\end{equation}

And we found that the spectra reconstructed by these 12 ICs excellently
match with the SSP database.

\subsection{Fitting Galaxy Spectra}\label{sec:fit}

The aim of this study is using these estimated 12 ICs to fit galaxy
spectra from large surveys.
The SFHs of galaxy can be approximated as a combination of
discrete bursts, thus the population of galaxy can be decomposed into
SSPs combination.  As shown in \S\,\ref{sec:ssica}, the SSP database
can be compressed into 12 ICs, so the model of observed galaxy spectra,
$\rm f_g(\lambda)$, can be fitted by these 12 ICs as:
\begin{equation}
{\rm f_g(\lambda)}=r(\lambda)\sum_{k=1}^{12} a_{k}{\rm IC}_{k}(\lambda,\sigma) ,
\end{equation}
where $r(\lambda)$ is the reddening term, describes the intrinsic
starlight reddening, can be modeled by the extinction law of Charlot
 \& Fall (2000).  ${\rm IC}_{k}(\lambda,\sigma)$ is
the $k-$th IC convolved with a Gaussian function.  The Gaussian width
$\sigma$ corresponds to the stellar velocity dispersion of a galaxy.
During the fitting process, we mask points around the prominent lines,
such as Balmer lines (\he, \hg, \hd, \hb, \ha) and strong forbidden lines
(\oii$\lambda$3727, \neiii$\lambda$3869, \oiii$\lambda\lambda$4959,5007,
\hei$\lambda$5876, \oi$\lambda$6300, \nii$\lambda\lambda$6548,6584,
and \sii$\lambda\lambda$6717, 6721).

After subtracting the modeled stellar population spectrum, emission
lines can be fitted with Gaussians simultaneously, similarly as Tremonti
et al. (2004): the forbidden lines (\oii, \oiii, \oi, \nii\, and \sii)
are set to have the same line width and velocity offset, likewise for
Balmer lines (\hg, \hd, \hb\, and \ha). By using the procedures above,
the observed galaxy spectra can be quickly recovered.

\section{Reliability of the fitting method}

\subsection{Simulations}

In this section, we analyze the simulated galaxy spectra to examine the
reliability of the MF-ICA method. All simulated spectra are generated
from 2007 version of BC03 stellar population synthesis code.  For sake
of simplicity, we parameterize each SFH of simulated galaxy in terms
of an underlying continuous model superimposed with random bursts on it
\citep{ka03}.  The spectral energy distribution (SED) at time
$t$ of a stellar population characterized by an exponentially declining
star formation law $\psi(t) \propto e^{-\gamma t}$   is given by:
\begin{equation}
F_{\lambda}(t)=\int_0^t \psi(t-t')S_{\lambda}(t',Z)dt' ,
\end{equation}
where $S_{\lambda}(t',Z)$ is the power
radiated by an SSP of age $t'$ and metallicity $Z$ per unit wavelength
per unit initial mass.

The added SFHs are described below:
\begin{enumerate}
\item The time when
a galaxy begins forming stars $t$ is distributed uniformly
between 0.1 and 13.5 Gyr. Star formation timescale $\gamma$ is uniformly
distributed between 0 and 1 $\rm{Gyr}^{-1}$.
\item Random bursts
occur at any time with same probability.  Bursts are parameterized in
terms of the fraction of stellar mass produced, which is logarithmically
distributed between 0.03 and 4, and their duration can vary between 0.03
and 0.3 Gyr.
\item The metallicities $Z$ are uniformly distributed
between $0.02\,\Zsun$ and $2\,\Zsun$, which represent the range of
stellar metallicities inferred from the spectra of $\sim2\times10^5$
SDSS galaxies.
\end{enumerate}

We apply our spectral analysis method on 500 simulated spectra over the
range $3322 - 9200$\AA. We also use the extinction law of Charlot \&
Fall (2000) to attenuate each spectrum, where the absorption optical
depth $\tau_{\rm V}$ is uniformly distributed between 0 and 5. The
velocity dispersion $\sigma$ is distributed uniformly between 50 ${\rm
km\, s^{-1}}$ and 450 ${\rm km\, s^{-1}}$.  Finally we added Gaussian
noise with signal-to-noise ratio S/N $=10$, 20, 30, respectively.

\subsection{Results}
\begin{figure*} 
\centering
\includegraphics[angle=-0,width=0.8\textwidth]{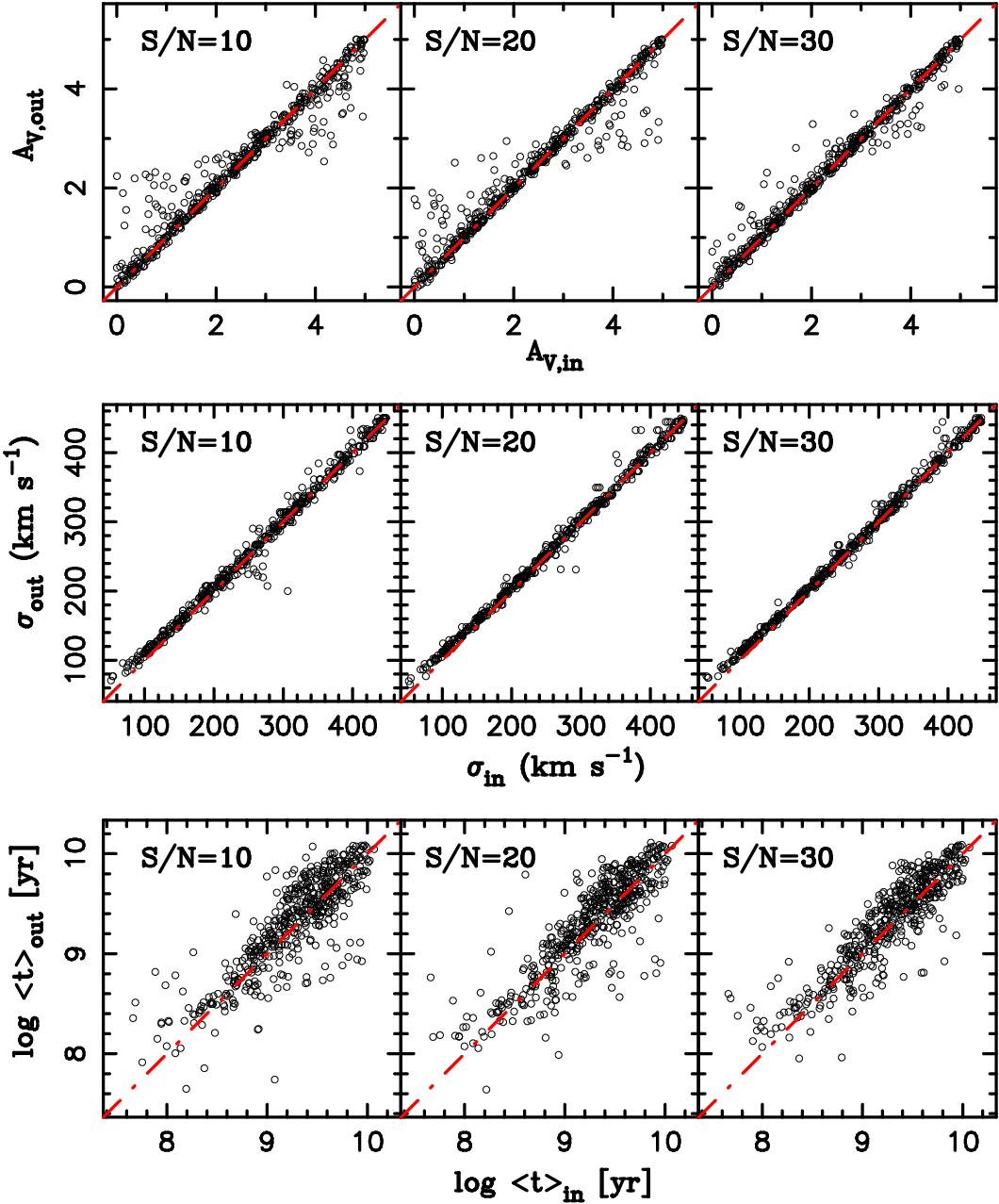}
\caption{
Comparison of the input $A_{\rm V}$ (magnitude), $\sigma$ (km s$^{-1}$)
and stellar ages (yr) with the output estimated values for simulations,
with S/N$=$10, 20 and 30, using our MF-ICA method.  
The red dot-dashed line is the identity line ($y=x$).}
\label{fig:sim}
\end{figure*}

From the fitting of simulated spectra, we expect to
examine the reliability of our spectral analysis approach which based
on MF-ICA algorithm. Our main parameters of interest are $\rm{A_{\rm
V}}$, $\sigma$, $t$ and $Z$.  Following steps are used to estimate age
and metallicity:
\begin{enumerate}
\item The pure spectrum of stellar
system of a galaxy, $f_g(\lambda)$, can be recovered by ICs, and it
also can be represented by a combination of $N$ SSPs.  Thus we can
solve the equation:
\begin{equation}
f_g(\lambda)=\sum_{k=1}^{12}a_{k}
{\rm IC_{k}(\lambda)}={\rm \sum_{j=1}^N} b_{j}{\rm f^j_{SSP}(\lambda)}.
 \label{eq:gl}
\end{equation}
\item We adopt 60 SSPs from CB07 database include models
of 15 different ages ($t=0.001$, 0.003, 0.005, 0.01, 0.025, 0.04, 0.1,
0.2, 0.6, 0.9, 1.4, 2.5, 5, 11, 13 Gyr) and 4 different metallicities
($Z=0.004$, 0.008, 0.02, 0.05).
\item After solving Eq. (\ref{eq:gl}),
the age and metallicity
 can be solved by:
\begin{equation}
{\rm \langle log \  t \rangle_{L}=\sum_{j=1}^{60}} b_j {\rm log(t_{j})} 
\end{equation}
\begin{equation} 
{\rm log \langle
\rm{Z} \rangle_{L}=\rm{log} \sum_{j=1}^{60} } b_j{\rm Z_{j}}.
\end{equation}
\end{enumerate}
Figure~\ref{fig:sim} shows the input parameters against estimated values
from simulated spectra with S/N $=10$, 20, 30. Clearly, the values of
starlight reddening $A_{\rm V}$ and stellar velocity dispersion $\sigma$
are almost well recovered. The mean square errors (MSE) between recovered
and input values are less than 0.20 and 7.45, respectively, as shown in
Table 1.

From the above method, a galaxy spectrum can be decomposed of 60
SSPs with weights.  The estimated weights $b_{j}$ can reflect the
fractional contributions of $j-$th SSP with age  ${\rm t_{j}}$ and metallicity
${\rm Z_{j}}$. Therefore, the light-weighted age and metallicity can be
estimated.  As shown in Figure~\ref{fig:sim} (bottom panel),  the
recovered and input values of $\langle \rm{log}\  t \rangle_{L}$ have
no significant difference with MSE less than 0.20.
According to the age-metallicity degeneracy problem \citep{br96}, 
the values of $\rm{log}\langle \rm{Z} \rangle_{L}$ recovered by our method 
cannot be exactly accurate. However, we can estimate them with meaningful 
accuracy, the Spearman's rank correlation coefficient $r_{\rm s}$ between 
output and input $\rm{log} \langle Z \rangle_{L}$ is about 0.70 for S/N $=20$.  
Finally, the summary of mean square error of parameters from simulated spectra 
can be found in Table 1.

\subsection{Compare to EL-ICA method}\label{sec:com}

\begin{figure*}
\centering
\includegraphics[angle=-90,width=0.9\textwidth]{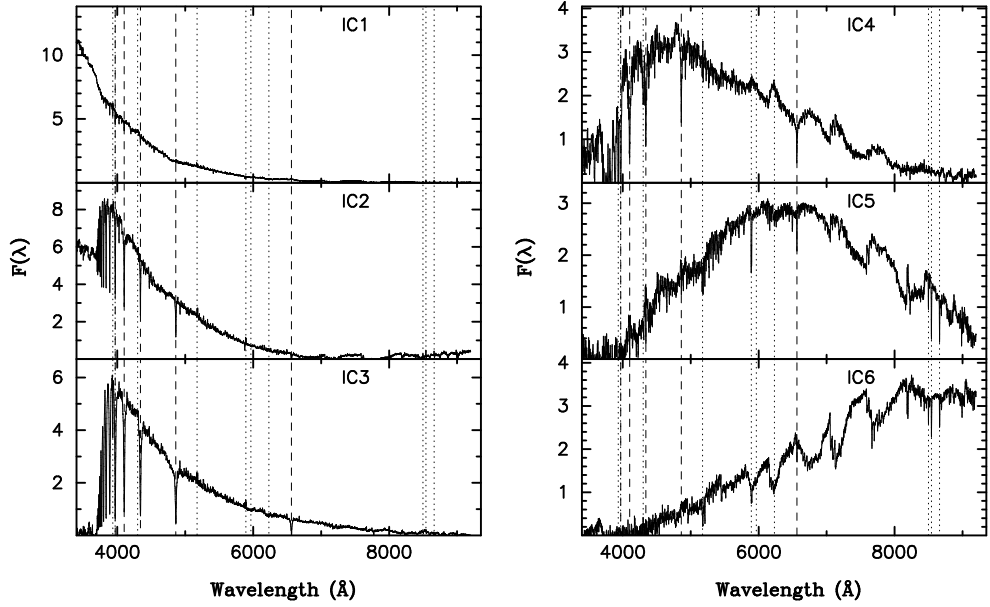}
\caption{
The spectra of 6 ICs estimated by EL-ICA method, some prominent spectral
features are labeled same as Figure\,4 in Su et al. (2013)}.
\label{fig:elica}
\end{figure*}

To carefully test the influence of ICA algorithms, we re-estimated
the ICs by EL-ICA algorithm \citep{mm01}.
The EL-ICA method, which also known as naive
mean field ICA method, has been applied in galaxy spectrum analysis
by Lu \etal (2006).  Here we used the same steps as Lu \etal (2006) and
also derived six ICs, we present these six ICs in Figure~\ref{fig:elica}.

\begin{figure*}
\centering
\includegraphics[angle=-90,width=0.9\textwidth]{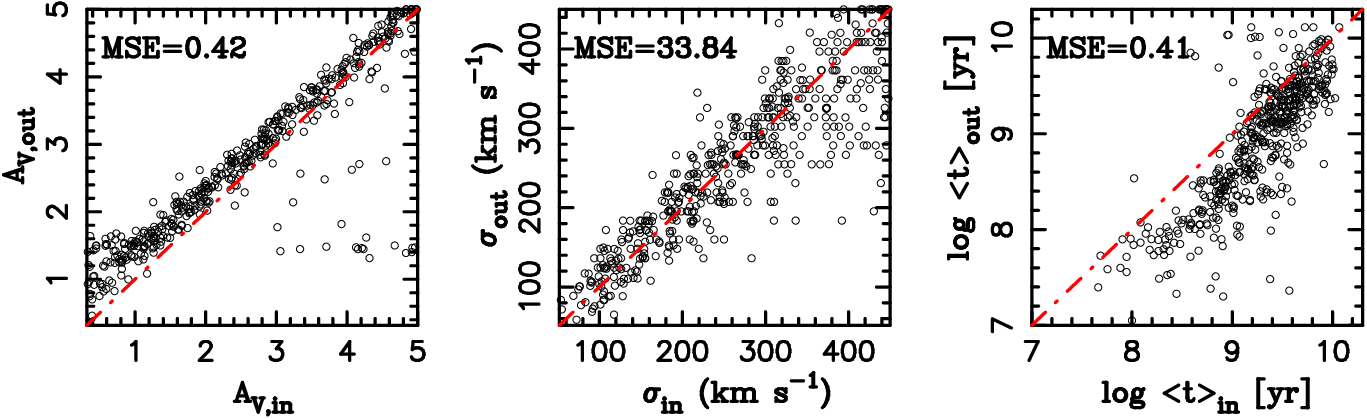}
\caption{
Comparison of the input $A_{\rm V}$ (magnitude), $\sigma$ (km s$^{-1}$) 
and stellar ages (yr) with the output estimated values for simulations, 
with S/N$=$20, using the EL-ICA method.  
The red dot-dashed line is the identity line ($y=x$).}
\label{fig:elsim}
\end{figure*}

We use these ICs to refit the simulated spectra, the input parameters
against output estimated values and the mean square error between
them for simulations with S/N=20 are shown in Figure~\ref{fig:elsim}.
The dispersions of parameters derived by EL-ICA method are larger than
those by our method, which were shown in Figure~\ref{fig:sim}.  And the
mean square errors of starlight reddening, velocity dispersion,
stellar age, metallicity ($A_{\rm V}$, $\sigma$, $\rm{\langle log\
t \rangle _{L}}$, $\rm{log\langle Z \rangle_{L}}$) are 0.421, 33.841,
0.405, and 0.299 for S/N=20, respectively.
These values are much larger than those of our method, as shown in
Table 1.
Finally, we also fit the SSP database using these ICs, the
spectra recovered are not as good as those by our method.
We conclude that our method which based on MF-ICA algorithm is more
precise and reliable.

\section{Application to SDSS spectra}

Using our MF-ICA spectral analysis method, we fit the SDSS galaxy
spectra, analyze stellar population properties of them, and measure
their emission-line properties from the starlight-subtracted spectra.
In this section, we compare the physical parameters obtained from
stellar population analysis of continuum and measurements of emission lines.
We also compare parameters estimated from our fitting technique with 
those derived by the MPA/JHU group.
Because the aim of this section is to test whether the results from our
spectral analysis method are reasonable and meaningful, we would not
investigate their physics properties, such as the formation and
evolution, of galaxies.

\subsection{Data preparation}\label{sec:data}

The Sloan Digital Sky Survey \citep{yo00} has released huge amounts of 
high-quality observed spectra of objects.  In this work, our spectra 
sample were extracted from spectroscopic plates of SDSS Data Release 8 
\citep{ai11}.  Moreover, we choose the objects which have been
spectroscopically classified as galaxies.  The spectra obtained from
SDSS span a wavelength from 3800\AA\ to 9200\AA\, with mean spectral
resolution $R=\lambda/\Delta\lambda \sim 1800$, and taken within three
arcsecond diameter fibers.
We finally fit about one million spectral sample of galaxies with
redshift less than 1, which obtained from SDSS spectroscopic
pipeline.

\begin{figure*} 
\centering
\includegraphics[angle=-90,width=0.9\textwidth]{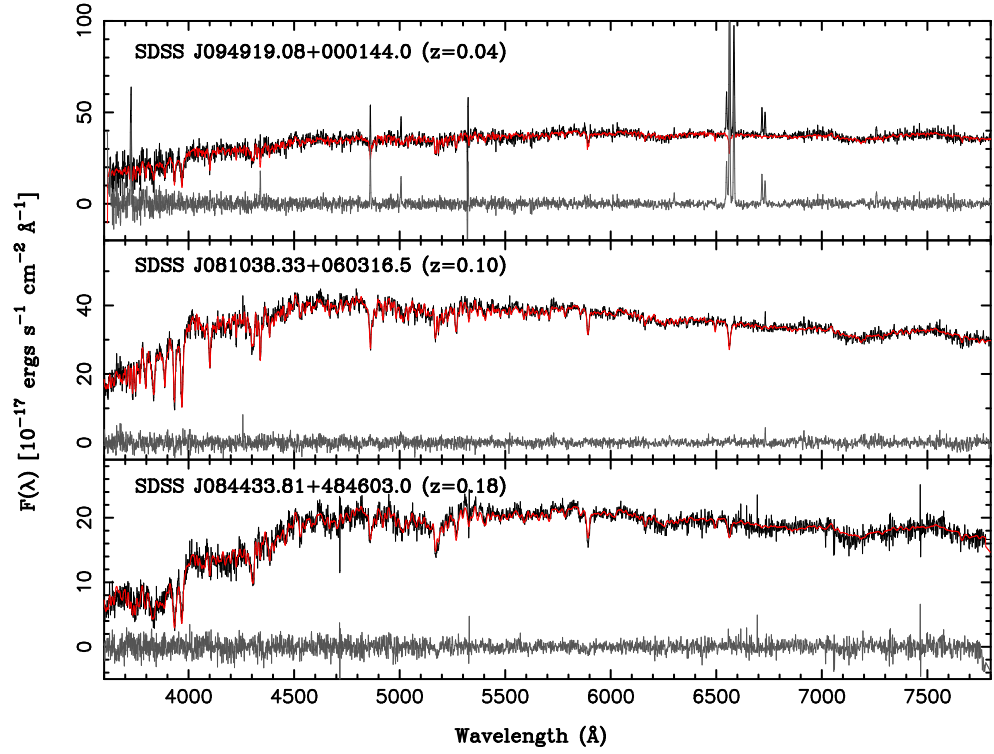}
\caption{
The spectra fitting results of some galaxies in our SDSS DR8 sample 
at a range of redshifts.  The black line shows the observed spectrum,
red line shows the modelled stellar spectrum, grey line shows the
residual spectrum, and the redshift is labeled in the top left 
corner of each panel.}
\label{fig:fit}
\end{figure*}

We use the MF-ICA method, which was described in \S\,\ref{sec:fit}, to fit
the spectral sample of galaxies from SDSS.
Firstly,
the spectra of galaxies were corrected for the foreground Galactic
extinction, using the maps of Schlegel et al. (1998).
Then, they were transformed into the rest frame, with spectroscopic
redshifts.  The spectral fittings results give a median
$\chi^{2}/$d.o.f (degree of freedom) of 1.13, nearly to excellent
value of 1 as we expected.  Figure~\ref{fig:fit} shows some examples
of the fitting, the spectra can be well recovered by eye-inspection,
which suggest that our MF-ICA spectral analysis approach works well.

\subsection{Comparisons with the MPA/JHU database}

The MPA/JHU group has provided catalogs
of estimated physical parameters of SDSS galaxies
publicly available on the website.\footnote{See
http://www.sdss3.org/dr8/spectro/spectro\_access.php} They inferred the
SFHs of DR8 galaxies on the basis of CB07 models, which was similar
to our researches.  Here we compare our own estimated parameters, such
as the emission line measurement, and stellar population properties,
with those obtained from the MPA/JHU catalogs.  Although we do not
expect our estimated parameters that perfectly consistent with them,
we analyze the relationships between these parameters to examine the
accuracy and reliability of MF-ICA algorithm.

\subsubsection{Stellar extinction}

\begin{figure*} \centering
\includegraphics[angle=-90,width=0.9\textwidth]{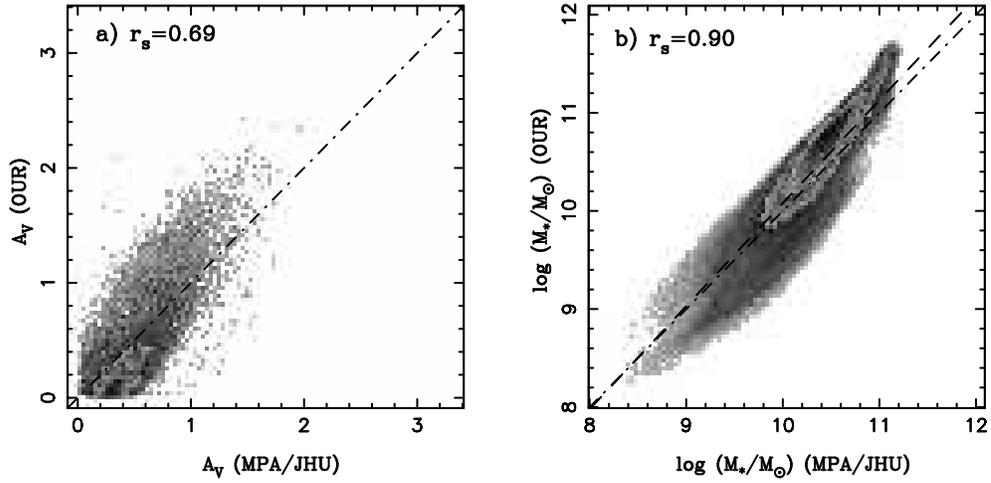}
\caption{
The values of stellar extinction $A_{\rm V}$ and stellar mass $M_*$ computed
from the MPA/JHU database versus those values computed by our code. 
The dot-dashed line is the identity line ($y=x$). The dashed line in the 
right panel is a robust fit for the relation.  
The number in the top left corner is the Spearman rank correlation 
coefficient.}
\label{fig:com}
\end{figure*}

In our fitting technique, the extinction of optical galaxy spectra is
modelled as one free parameter $A_{\rm V}$.  In Figure\,\ref{fig:com}a),
we plot the values of $A_{\rm V}$ estimated by our method, versus those
estimated by the MPA/JHU group, which adopt the same attenuation curve
by Charlot \& Fall (2000). Since only few galaxies with $A_{\rm V}<0$ 
are found in previous research works, we constrain $A_{\rm V}$
to be positive in our analysis.  Therefore, this constrains will not
have significant impact on the results of our analysis.

We adopt the value of Spearman's rank correlation coefficient $r_{\rm s}$
to describe the relationship between two variables.  As shown in Figure
\,\ref{fig:com}a), our results are well and linearly correlated to those
extracted from the MPA/JHU catalogs, with $r_{\rm s}=0.69$.  However, the
extinction values $A_{\rm V}$ which obtained from the MPA/JHU database
are systematically lower with our values, similar as found in Chen
et al. (2012, in Figure 3f).  One possible reason for this discrepancy
is that we only use the optical-band spectra to estimate the stellar
extinction $A_{\rm V}$.

\subsubsection{Stellar mass}

By using our stellar population analysis method, the light-weighted
stellar mass $\rm{log}\langle \rm{M} \rangle_{L}$ of SDSS galaxy also can
be estimated. We calculate M/L ratio by weighed-added M/L ratios of 
each SSP component, and then derive the stellar mass by multiplying it 
by luminosity. In Figure~\ref{fig:com}b), we plot our estimate stellar
mass against the MPA/JHU extinction-corrected stellar mass. The results
from two methods are well consistent, with $r_{\rm s}=0.90$.  The small
discrepancy is caused due to the different estimate methods.  
In our method, the stellar masses are obtained from the $M/L$ ratio, which 
estimated through the best $\chi^{2}$ model.  While the MPA/JHU group 
estimated their $M/L$ ratio through Bayesian inference method, which connect 
two indices, \hda\ and \dfn\,, with a model obtained from a large library
of Monte Carlo realizations of galaxies with different SFHs \citep{ka03}.

\subsubsection{Emission lines and nebular metallicities}

\begin{figure*}
\centering
\includegraphics[angle=-90,width=0.9\textwidth]{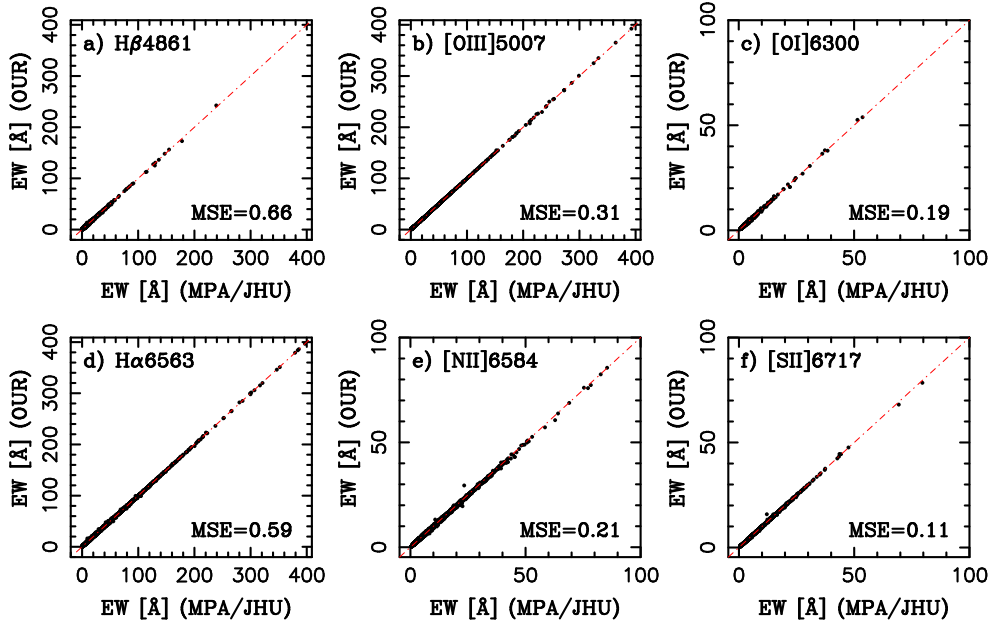}
\caption{
The comparison of equivalent widths of \hb, \oiii$\lambda$5007,
\oi$\lambda$6300, \ha, \nii$\lambda$6584 and \sii$\lambda$6717 measured
by the MPA/JHU group with those by our code. The red dot-dashed line is
the identity line ($y=x$), while the number in the bottom-right corner
of each panel indicates the mean square error (MSE).}
\label{fig:eml}
\end{figure*}

In our case the emission lines were measured from starlight-subtracted
spectrum.  The MPA/JHU group adopted a similar method, however, they
only used a single metallicity CB07 model to fit the observed continuum.
We plot our estimated equivalent widths (EWs) of emission lines, such
as \hb, \oiii$\lambda$5007, \oi$\lambda$6300, \ha, \nii$\lambda$6584
and \sii$\lambda$6717 versus those measured by the MPA/JHU group in
Figure\,\ref{fig:eml}.  
As shown in Figure\,\ref{fig:eml}, our values are consistent with those 
measured by the MPA/JHU group, with small discrepancy.
We adopt the MSE to quantify the discrepancy between them.  On the whole,
the MSE of all these values are less than 1, suggest that there are no
significant difference.  The small discrepancy appearing is due
to different measurement of the synthesized spectrum, which related to
different subtracted stellar absorption.

\begin{figure} 
\centering
\includegraphics[angle=-90,width=0.9\columnwidth]{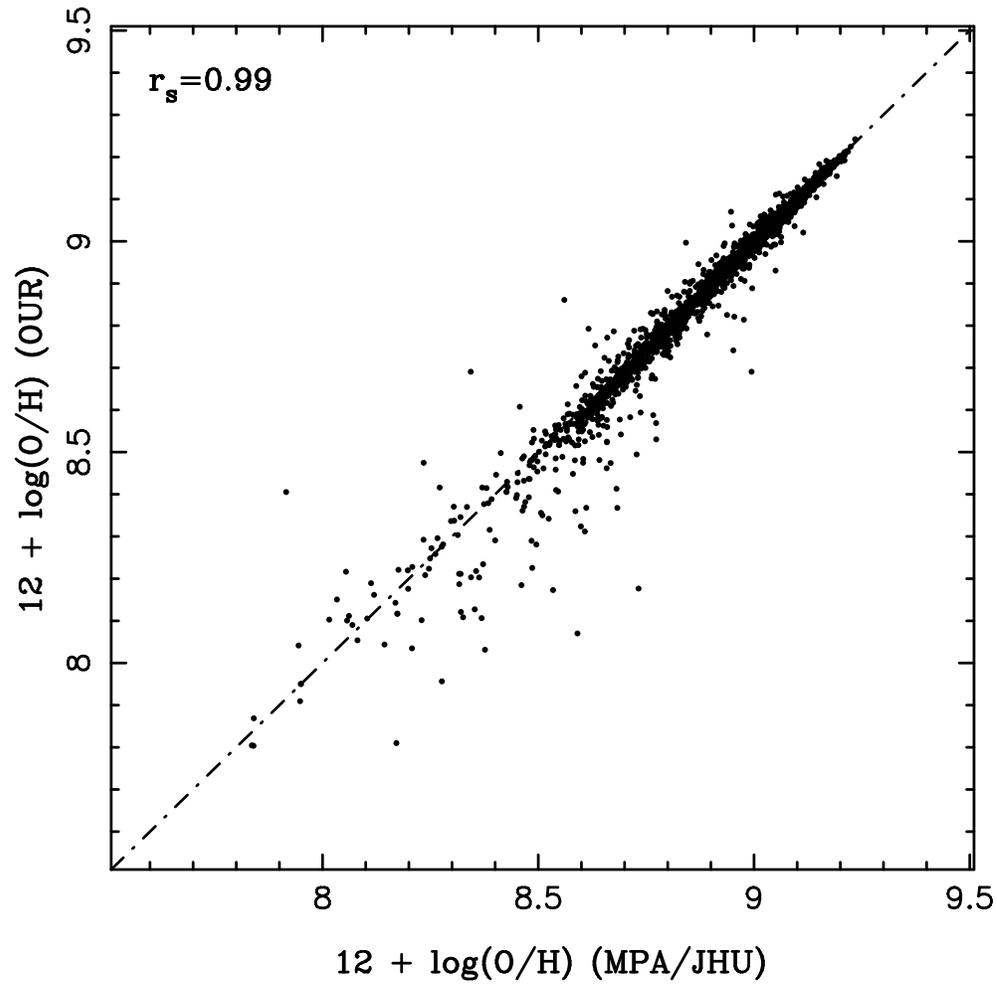}
 \caption{
Plot of our estimated nebular oxygen abundance against those obtained by 
the MPA/JHU group.  The dot-dashed line is the identity line ($y=x$), 
while the number in the top left corner is the Spearman rank correlation 
coefficient.}
\label{fig:z}
\end{figure}

We also compared the value of nebular oxygen abundance \zoh, which
can be obtained from the equation described in Tremoni \etal (2004).
As shown in Figure\,\ref{fig:z}, our estimated values of nebular oxygen
abundance show a high degree of correspondence with those drawn from
the MPA/JHU catalog.  The value of Spearman rank coefficient is 0.99,
nearly to a perfect Spearman correlation of $r_{\rm s}=1$, which
indicates the perfect monotonic relationship.

This part is summarized as follows.
We have compared estimated parameters such as the stellar extinction,
stellar mass and emission line measurement with those obtained from
the MPA/JHU catalogs.  According to the analysis of relationships between
these parameters, we conclude that MF-ICA method is reasonable and 
reliable.

\subsection{Empirical relations}

In this subsection, another way was used to test the accuracy of our
method. The parameters estimated from the analysis of continuum were 
compared with those estimated by
using measured emission lines.  We analyze the relationships between
these parameters to investigate whether our method derived results
are reasonable.

\subsubsection{Relations between Balmer features and stellar age}

\begin{figure*}
\centering
\includegraphics[angle=-90,width=0.9\textwidth]{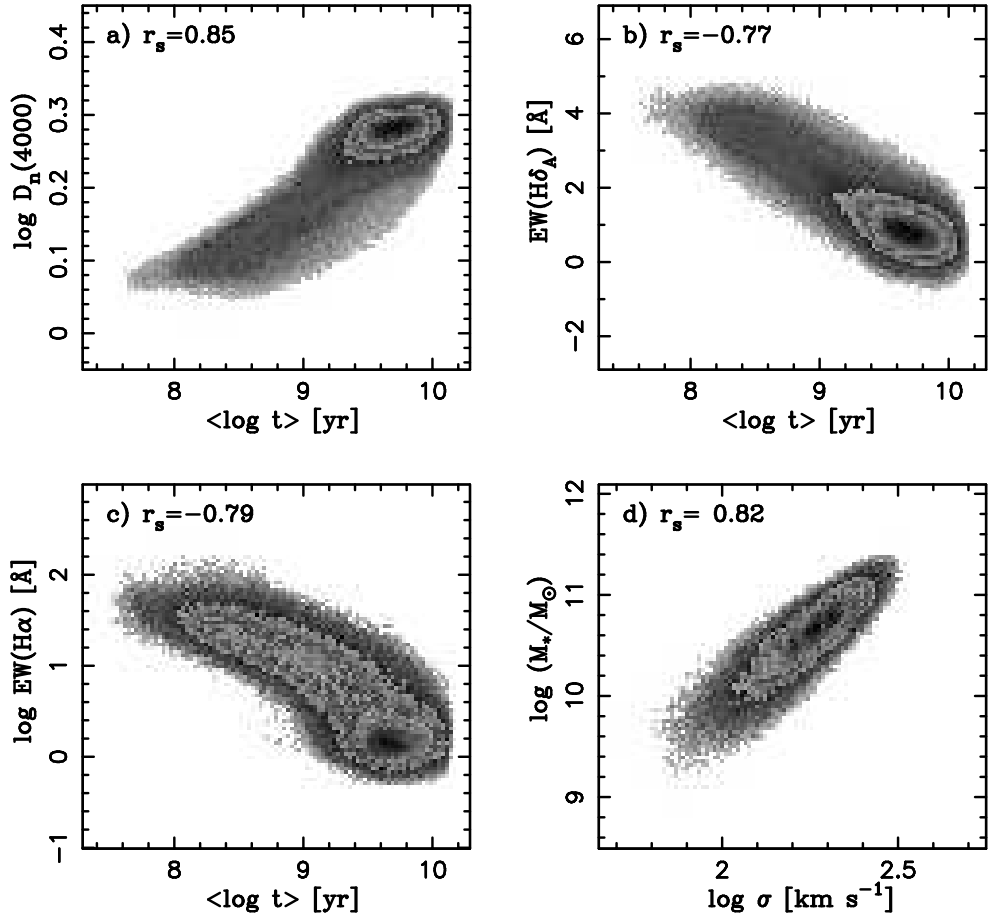}
\caption{ 
Relations of the 4000\AA\, break index versus the light-weighted mean stellar
 age (a);
the \hda\, index versus the light-weighted mean stellar age (b); the
equivalent width of \ha\, versus the light-weighted mean stellar age (c);
the comparison of our estimated stellar mass (log M$_*$) with velocity
dispersion (log $\sigma$) (d).  
The number in the top left corner of each panel is
the Spearman rank correlation coefficient $r_{\rm s}$.}
\label{fig:rel}
\end{figure*}

The value of 4000 \AA\, break index \citep{ba99} can reflect
age of galaxy.  Higher \dfn\, values relates to older, metal--rich
galaxies, while lower values related to younger stellar subpopulation
of galaxies.  The strength of \hda\, \citep{wo97} is
another age indicator. Strong \hda\ absorption of galaxy reflects a
burst of star formation occurred in the past $0 - 1$ Gyr.  Therefore,
our estimated ages of galaxies should be increased with \dfn\ values,
decreased with \hda\ values.  In Figure\,\ref{fig:rel}a) and b),
correlations between ages, \dfn\, and \hda\ values are shown expected
relationships, $\dfn - \langle \rm{log \ t} \rangle_{L}$ trends with
$r_{\rm s}= 0.85$ are strongly positive, and $\hda - \langle \rm{log \
t} \rangle_{L}$ trends with $r_{\rm s}= -0.77$ are strongly negative.

For the galaxy with emission lines, \ha\ emission--line is corresponded to
the instantaneous star formation rate (SFR) of a galaxy 
\citep{2012ARA&A..50..531K}.
Therefore the equivalent width (EW) of \ha\, is also an age indicator,
which would be lager for younger galaxies.  Figure\,\ref{fig:rel}c) shows
that the light-weighted stellar age $\langle \rm{log \ t} \rangle_{L}$
correlate negatively with EW(\ha) ($r_{\rm s} = -0.79$), as we expected.
These relations respect that the stellar ages $\langle \rm{log \ t}
\rangle_{L}$ we obtained by our spectral synthesis are reasonable.

\subsubsection{Stellar mass and velocity dispersion}

According to the viral theorem, for constant mass surface density,
the stellar mass (log $M_*$) is expected to be correlated positively with
the stellar velocity dispersion (log $\sigma$).  In the sample of old
galaxies, $\sigma$ is related to galaxy mass through the Faber-Jackson
relation.  Moreover, the stellar velocity dispersion of young, star
forming galaxies is contributed from bulge and disc, thus it is related to
galaxy mass through the Tully-Fisher relation.
In Figure\,\ref{fig:rel}d), we plot our estimates of stellar mass (log $M_*$) 
with velocity dispersion (log $\sigma$).  The $M_*-\sigma$ relation 
trend strongly positive with $r_{\rm s} =0.82$ as we expected, which 
suggests our synthesis results are meaningful.

We have analyzed the correlations between physical parameters obtained
from stellar populations, such as stellar ages and stellar masses, with
independent quantities.  The strong correlations of $\rm{\langle log\
t \rangle_{L}} - \dfn$, \hda, EW(\ha), and M$_* - \sigma$ suggest
that the parameters derived by our spectral synthesis approach through
MF-ICA method are reasonable and meaningful.

\section{Application to spectra of galaxies with higher redshift}

Optical galaxy redshift surveys are not only vital importance in
cosmology, but also very important to understand physical processes
related to galaxy formation and/or evolution\citep{2015RAA....15..819F}.  
In the last few years,
redshift surveys, such as 2dF Galaxy Redshift Survey \citep{co01} and SDSS, 
have measured redshifts of millions of low redshift (with median
 redshift of $z=0.1$) 
galaxies. With larger aperture telescope, new generation redshift surveys, 
such as
the DEEP2, BigBOSS, LAMOST, will measure redshifts of galaxies with higher
redshift \citep{da07,sc09,ks10,2011RAA....11.1093Z}.
The motivation for this work is that we want to provide an easy-to-use
full-spectrum fitting package and determine spectral parameters for
spectra of the LAMOST extragalactic surveys \citep{ks10}.  Since the
regular spectroscopic survey of LAMOST just beginning, we apply our
synthesis approach to the spectra of galaxy from the DEEP2 survey,
which has a similar signal-to-noise ratio as the spectrum of LAMOST 
\citep{2015arXiv150501570L}.

\begin{figure*} 
\centering
\includegraphics[angle=-90,width=0.9\textwidth]{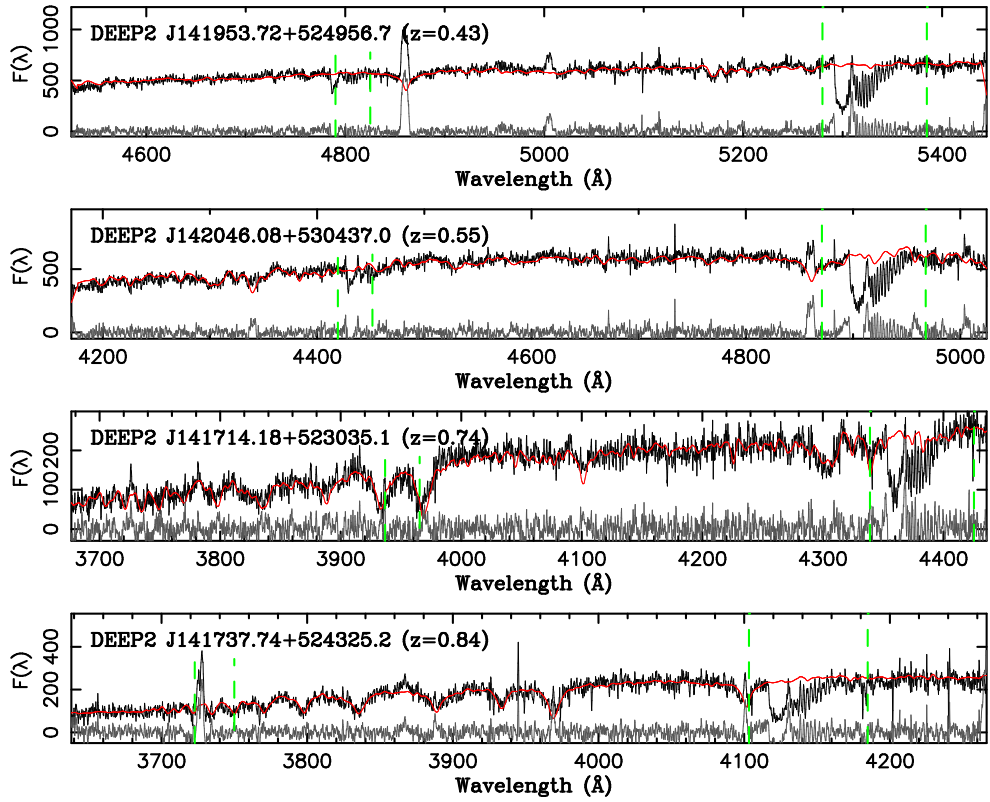}
\caption{
The spectra fitting results of some galaxies in our DEEP2 sample at
a range of redshifts, which are labeled in the top left cornor of 
each panel.  The black line shows the observed spectrum,
red line shows the modelled stellar spectrum, and grey line shows the
residual spectrum. We also mask the ``telluric absorption''  regions
between dashed lines (observed frame: $7750 - 7700$\AA\, \& $6850 -
6900$\AA).}
\label{fig:d2}
\end{figure*}

In the Extended Groth Strip (EGS) field, utilizing the DEIMOS (Deep
Imaging Muti-object spectrograph) boarded on the Keck 10 m telescope,
DEEP2 galaxy redshift survey provides the spectral data of galaxies with
the redshift from 0 to 1.4.  DEIMOS have highest-revolution grating, 1200
line mm$^{-1}$, typically covering $6500 - 9100$\AA, with a spectral
resolution $R=\lambda/\Delta\lambda \sim 6000$ \citep{2003SPIE.4841.1657F}.
In our study, we only analyze galaxies with redshift quality $Q\geq3$,
Thus, we obtained 9501 galaxies with $Q\geq3$ in the EGS, corresponding
to the median redshift is 0.74.  The details of these galaxies spectra
extraction can be found in Davis et al. (2007). Finally, we obtained
about 1,400 sources with $\rm{S/N} > 3$, and show some examples of the
fitting in Figure\,\ref{fig:d2}.  It can be seen that our MF-ICA fitting
method work well, and we will analysis their physical properties in the
future work.

\section{Summary}

In this work, we have presented the MF-ICA method to compress the
CB07 SSP library into a few Independent Components (ICs)
as templates to fit observed galaxy spectra.  Although there are many
statistical multivariate data processing techniques, MF-ICA seems to be
more useful.  Since it has the capability of providing good estimates of
the results by selecting proper parameters.  The goal of our project was
to estimate physical properties quickly and accurately for a large sample
of galaxies.  By using MF-ICA algorithm, we can fit an observed spectrum
of galaxy only a few second, which is time-efficient for analysis of
galaxies spectra observed by large-area surveys, such as LAMOST, BigBOSS.

We have tested our method to fit the simulated and the SDSS DR8 galaxy
spectra. Simulations show that the important parameters of galaxies can
be accurately recovered by our method, such as stellar contents, star
formation histories, starlight reddening and stellar velocity dispersion.

We have compared parameters estimated from our fitting technique to
those obtained from the MPA/JHU group of DR8 galaxies.  These physical
parameters and measurements are in good agreement.  We also analyze the
correlations between physical parameters obtained from stellar populations
with independent quantities.  We find strong correlations of M$_* -
\sigma$, $\rm{\langle log\ t \rangle_{L}} - \dfn$, \hda , EW(\ha).

In future studies, we intent to apply our fitting technique to other
large data bases, such as the LAMOST ExtraGAlactic Surveys (LEGAS),
the DEEP2 galaxy redshift survey.  We have fitted more than 1,400 DEEP2
galaxies spectra, our next step will analysis their physical properties.

\section*{Acknowledgments} 
We are grateful to the anonymous referee for making constructive 
suggestions to improve the paper.
We thank Stephane Charlot for providing the unpublished CB07 stellar 
population synthesis models and helpful discussions. 
This work is supported by the National Natural Science Foundation of 
China (NSFC, Nos. 11225315, 1320101002, 11433005, and 11421303), the 
Strategic Priority Research Program "The Emergence of Cosmological 
Structures" of the Chinese Academy of Sciences (No. XDB09000000), the 
Specialized Research Fund for the Doctoral Program of Higher Education
(SRFDP, No. 20123402110037), and the Chinese National 973 Fundamental 
Science Programs (973 program) (2015CB857004).

\clearpage

\begin{table}
\begin{center}
\begin{minipage}[]{100mm}
\caption{ Summary of Parameter Error Estimates for Simulated Spectra. 
The different rows list the mean square error (MSE) between 
output and input values of the corresponding quantity, as obtained from 
simulations with different signal-to-noise ratios (S/N).}
\end{minipage}
\begin{tabular}{lrrrr}
\hline\noalign{\smallskip}
S/N & $\rm{MSE}_{Av}$ (mag) & $\rm{MSE}_\sigma$ (km s$^{-1}$) & $\rm{MSE}_{\rm{\langle 
log\ t\rangle _{L}}}$ & $\rm{MSE}_{\rm{log\langle Z \rangle_{L}}}$ \\
\hline\noalign{\smallskip}
10 	& 0.191  & 7.449            & 0.201     &0.201\\
20 	& 0.169  & 6.301            & 0.189     &0.202\\
30      & 0.119  & 6.017           & 0.169      &0.196\\
\noalign{\smallskip}\hline
\end{tabular}
\end{center}
\end{table}

\label{lastpage}

\end{document}